\documentclass[12pt]{article}
\usepackage{graphicx}
\usepackage{color}
\newcommand{\xrm}[1]{{\textstyle \mbox{\rm #1}}}
\newcommand{\fnd}[2]{\frac{\textstyle #1}{\textstyle #2}}
\newcommand{\fndrs}[4]{\fnd{\raisebox{#1}{$#2$}}{\raisebox{#3}{$#4$}}}
\newcommand{\dissum}[2]{\displaystyle \sum_{#1}^{#2}}
\newcommand{\disprod}[2]{\displaystyle \prod_{#1}^{#2}}
\newcommand{\bm}[1]{\mbox{\boldmath $#1$}}
\newcommand{\x}[1]{{\textstyle #1}}
\newcommand{\bracket}[2]{\mbox{$\left\langle #1\left| #2\right.\right\rangle$}}
\newcommand{\braket}[3]{\mbox{$\left\langle #1\left|
#2\right| #3\right\rangle$}}
\def\chie{\mbox{\raisebox{0.5ex}{$\chi$}}}
\newcommand{\Clebsch}[6]{\mbox{$\left(\begin{array}{rrr}
#1 & #2 & #3 \\ [3pt] #4 & #5 & #6 \end{array}\right)$}}
\begin{document} \baselineskip .7cm
\title{\bf Recoupling matrix elements and decay}

\author{
E. van Beveren$^{1}$,\\
{\normalsize\it Institute for Theoretical Physics, University of Nijmegen}\\
{\normalsize\it NL-6525 ED, Nijmegen, The Netherlands}
}
\footnotetext[1]{present address:
Departamento de F\'{\i}sica, Universidade de Coimbra, P-3004-516 Portugal\\
e-mail: eef@teor.fis.uc.pt\\
url: http://cft.fis.uc.pt/eef}

\date{preprint, 21 July 1982\\ - \\
published in \\ Zeitschrift f\"{u}r Physik {\bf C} - Particles and Fields
{\bf 17}, 135-140 (1983)}
\maketitle

\begin{abstract}
Recoupling matrix elements are evaluated, in the harmonic oscillator
approximation, for all possible angular and radial excitations in processes
where quarks recombine.
A diagrammatic representation is given.
Their use is demonstrated in calculating the transition potential
for $\rho\to 2\pi$ in a pair creation model.
\end{abstract}
\clearpage

\section{Introduction}

The description of decaying particles has already a long history.
Nevertheless, only recently phenomenological potential models
\cite{AP105p318,PRD17p3090,PRD21p772}
have been proposed, for which the description of wave functions,
hadronic widths, resonance scattering, ..., goes beyond
perturbative methods.

These potential models are based on the $^{3}P_{0}$ mechanism for decay.
It is therefore necessary to have a scheme along which
the coupling constants for the coupling
of many quark channels to many hadron channels can easily be calculated.
In this work we propose a scheme which is based on ideas which are
intimately connected with the potential model of \cite{PRD21p772}.

We assume that, during the process of the decay of
a hadron, an intermediate situation occurs in which the original quarks
and the newly formed $^{3}P_{0}$ pair are kept together in a kind
of hadronic soup.
The formation of new hadrons out of this soup is then assumed to be caused
by rearrangements of the quarks.

The forces in the intermediate soup are taken
to be harmonic oscillator forces.
This has two advantages:
\begin{itemize}
\item
the number of possible final states is limited;
\item
the calculations can easily be performed.
\end{itemize}

Since the complications of spin are easy to handle in this scheme,
because spin is not connected with space coordinates,
we confine ourselves to the treatment of the space part of the
so-called recoupling matrix elements$^{2}$.
\footnotetext[2]{Spin, isospin and color degrees of freedom
have in the mean time been dealt with in a separate paper:\\
E.~van Beveren,
{\it Coupling constants and transition potentials for hadronic decay
modes of a meson},\\
Z.\ Phys.\ C {\bf 21}, 291 (1984), [arXiv:hep-ph/0602246].}

In Sect.~\ref{MtoMM} we introduce as an example explicitly
an orthogonal transformation which describes the rearrangement
of quarks and antiquarks for the case
of the decay of a meson into two mesons.
The calculation of the recoupling matrix elements for the general case
is performed in Sect.~\ref{DME}.
The central result of this paper is given in formula (\ref{Dfinal})
and by the related diagrammatic representation which is shown
in Fig.~\ref{diagrep}.
In Sect.~\ref{MtoMMTpot} we show how, within this formalism, we can calculate
the spatial dependence of the transition potential for the decay
of a $\rho$ meson into two pions.
\clearpage

\section{The decay of a meson into two mesons}
\label{MtoMM}

After the formation of an extra quark-antiquark pair out of the vacuum
($^{3}P_{0}$) in a decaying meson,
we can treat this meson as a system of four particles.
We have for instance, quarks 1 and 3 and antiquarks 2 and 4.
Suppose that 1 and 4 come from the decaying meson
and 2 and 3 from the $^{3}P_{0}$ pair,
then we study the coupling of this system to a meson-meson final state
where the flavor content of one meson consists of flavors 1 and 2
and of the other meson of flavors 3 and 4.

We thus study a system of four particles with constituent masses
$\mu_{1}$, $\mu_{2}$, $\mu_{3}$ and $\mu_{4}$.
The positions and the momenta of the quarks and antiquarks are given by
\bm{r}$_{i}$ and \bm{p}$_{i}$ ($i=$ 1,...,4) respectively.
On grounds of our investigations \cite{THEFNYM7911} it is reasonable
to place the particles 1 to 4 in independent mass-dependent
non-relativistic harmonic oscillators.
The oscillator frequency $\omega$ is universal as is noticed in
\cite{PRD21p772,THEFNYM7911}.
In the centre of mass system the motion of the four particles
is described by the Hamiltonian:
\begin{equation}
{\cal H}\; =\;
\sum_{i=1}^{4}\,
\fnd{\bm{p}_{i}^{2}+\left(\omega\mu_{i}\bm{r}_{i}\right)^{2}}
{2\mu_{i}}
\; -\;
\fnd{
\left(\dissum{i=1}{4}\,\bm{p}_{i}\right)^{2}+
\left(\dissum{i=1}{4}\,\omega\mu_{i}\bm{r}_{i}\right)^{2}
}
{2\left(\mu_{1}+\mu_{2}+\mu_{3}+\mu_{4}\right)}
\;\;\; .
\label{4pHamiltonian}
\end{equation}
The Hamiltonian (\ref{4pHamiltonian}) can be expressed in terms of
three independent dimensionless coordinates and momenta.
If we want to describe the situation before decay we may choose
the motion of 4 with respect to 1, the motion of 2 with respect to 3
and the motion of the pair (3 + 2) with respect to the pair (1 + 4).
For this we introduce dimensionless coordinates
\bm{\rho} and momenta \bm{\pi}:
\begin{eqnarray}
& &
\bm{\rho}_{ij}\; =\;\left(\bm{r}_{j}-\bm{r}_{i}\right)
\sqrt{\fnd{\omega\mu_{i}\mu_{j}}{\mu_{i}+\mu_{j}}}
\;\;\; ,\;\;\;
\bm{\pi}_{ij}\; =\;
\left(\fnd{\bm{p}_{j}}{\mu_{j}}-\fnd{\bm{p}_{i}}{\mu_{i}}\right)
\sqrt{\fnd{\omega^{-1}\mu_{i}\mu_{j}}{\mu_{i}+\mu_{j}}}
\nonumber\\ [10pt] & &
\bm{\rho}_{ijk\ell}\; =\;
\left(
\fnd{\mu_{k}\bm{r}_{k}+\mu_{\ell}\bm{r}_{\ell}}{\mu_{k}+\mu_{\ell}}-
\fnd{\mu_{i}\bm{r}_{i}+\mu_{j}\bm{r}_{j}}{\mu_{i}+\mu_{j}}
\right)
\sqrt{\fnd{\omega\left(\mu_{i}+\mu_{j}\right)\left(\mu_{k}+\mu_{\ell}\right)}
{\mu_{1}+\mu_{2}+\mu_{3}+\mu_{4}}}
\;\;\;\xrm{and}
\nonumber\\ [10pt] & &
\bm{\pi}_{ijk\ell}\; =\;
\left(
\fnd{\bm{p}_{k}+\bm{p}_{\ell}}{\mu_{k}+\mu_{\ell}}-
\fnd{\bm{p}_{i}+\bm{p}_{j}}{\mu_{i}+\mu_{j}}
\right)
\sqrt{
\fnd{\omega^{-1}\left(\mu_{i}+\mu_{j}\right)\left(\mu_{k}+\mu_{\ell}\right)}
{\mu_{1}+\mu_{2}+\mu_{3}+\mu_{4}}}
\;\;\; .
\label{rhoandpi}
\end{eqnarray}
The Hamiltonian which governs the situation before decay is
given by formula (\ref{4pHamiltonian})
expressed in terms of the coordinates and momenta
$\left(\bm{\rho}_{14},\bm{\pi}_{14}\right)$,
$\left(\bm{\rho}_{32},\bm{\pi}_{32}\right)$ and
$\left(\bm{\rho}_{1432},\bm{\pi}_{1432}\right)$:
\begin{equation}
{\cal H}\; =\;
\frac{1}{2}\omega\,\left\{
\bm{\pi}_{14}^{2}+
\bm{\pi}_{32}^{2}+
\bm{\pi}_{1432}^{2}+
\bm{\rho}_{14}^{2}+
\bm{\rho}_{32}^{2}+
\bm{\rho}_{1432}^{2}
\right\}
\;\;\; .
\label{4pHamiltonian1432}
\end{equation}
Analogously we can describe the situation after decay
at best with the help of the coordinates and momenta
$\left(\bm{\rho}_{12},\bm{\pi}_{12}\right)$,
$\left(\bm{\rho}_{34},\bm{\pi}_{34}\right)$ and
$\left(\bm{\rho}_{1234},\bm{\pi}_{1234}\right)$,
which respectively belong to the motion of 2 with respect to 1, of 4
with respect to 3 and of the pair (3 + 4) with respect to the pair (1 + 2),
{\it i.e.}
\begin{equation}
{\cal H}\; =\;
\frac{1}{2}\omega\,\left\{
\bm{\pi}_{12}^{2}+
\bm{\pi}_{34}^{2}+
\bm{\pi}_{1234}^{2}+
\bm{\rho}_{12}^{2}+
\bm{\rho}_{34}^{2}+
\bm{\rho}_{1234}^{2}
\right\}
\;\;\; .
\label{4pHamiltonian1234}
\end{equation}
It is an easy task to check that the transformation which transforms
(\ref{4pHamiltonian1432}) into (\ref{4pHamiltonian1234}) is given by
\begin{equation}
\alpha =
\fndrs{53pt}
{
\left(
\begin{array}{ccc}
\sqrt{\mu_{2}\left(\mu_{2}+\mu_{3}\right)\left(\mu_{3}+\mu_{4}\right)\mu_{4}} &
\;\;
\sqrt{\mu_{3}\left(\mu_{3}+\mu_{4}\right)\left(\mu_{4}+\mu_{1}\right)\mu_{1}}
\;\; &
\sqrt{\mu_{1}\mu_{2}\left(\mu_{3}+\mu_{4}\right)\dissum{i=1}{4}\mu_{i}}
\\ [14pt]
\sqrt{\mu_{1}\left(\mu_{1}+\mu_{2}\right)\left(\mu_{2}+\mu_{3}\right)\mu_{3}} &
\sqrt{\mu_{4}\left(\mu_{4}+\mu_{1}\right)\left(\mu_{1}+\mu_{2}\right)\mu_{2}} &
-\sqrt{\left(\mu_{1}+\mu_{2}\right)\mu_{3}\mu_{4}\dissum{i=1}{4}\mu_{i}}
\\ [10pt]
\sqrt{\mu_{1}\left(\mu_{2}+\mu_{3}\right)\mu_{4}\dissum{i=1}{4}\mu_{i}} &
-\sqrt{\left(\mu_{4}+\mu_{1}\right)\mu_{2}\mu_{3}\dissum{i=1}{4}\mu_{i}} &
\mu_{1}\mu_{3}-\mu_{2}\mu_{4}
\end{array}
\right)
}
{-8pt}
{
\sqrt{
\left(\mu_{1}+\mu_{2}\right)\left(\mu_{2}+\mu_{3}\right)
\left(\mu_{3}+\mu_{4}\right)\left(\mu_{4}+\mu_{1}\right)
}
}
\label{BAtrafo}
\end{equation}
For the matrix $\alpha$ it has to be understood that
\begin{equation}
\left(
\begin{array}{l}
\bm{\rho}_{12}\\ [10pt]
\bm{\rho}_{34}\\ [10pt]
\bm{\rho}_{1234}
\end{array}
\right)
\;\;\left\{\;\xrm{or}\;\;
\left(
\begin{array}{l}
\bm{\pi}_{12}\\ [10pt]
\bm{\pi}_{34}\\ [10pt]
\bm{\pi}_{1234}
\end{array}
\right)
\right\}\; =\;
\left(
\begin{array}{ccc}
\alpha_{11} & \alpha_{12} & \alpha_{13}\\ [10pt]
\alpha_{21} & \alpha_{22} & \alpha_{23}\\ [10pt]
\alpha_{31} & \alpha_{32} & \alpha_{33}
\end{array}
\right)
\left(
\begin{array}{l}
\bm{\rho}_{14}\\ [10pt]
\bm{\rho}_{32}\\ [10pt]
\bm{\rho}_{1432}
\end{array}
\right)
\;\;\left\{\;\xrm{or}\;\;
\left(
\begin{array}{c}
\bm{\pi}_{14}\\ [10pt]
\bm{\pi}_{32}\\ [10pt]
\bm{\pi}_{1432}
\end{array}
\right)
\right\}
\;\;\; .
\label{alphamatrix}
\end{equation}
The transformation $\alpha$ is orthogonal.

For the volume element which determines the normalization
of the wave functions, we choose
\begin{eqnarray}
& &
d^{3}\left(\bm{r}_{2}-\bm{r}_{1}\right)\,
d^{3}\left(\bm{r}_{4}-\bm{r}_{3}\right)\,
d^{3}\left(
\fnd{\mu_{3}\bm{r}_{3}+\mu_{4}\bm{r}_{4}}{\mu_{3}+\mu_{4}}-
\fnd{\mu_{1}\bm{r}_{1}+\mu_{2}\bm{r}_{2}}{\mu_{1}+\mu_{2}}
\right)\; =\nonumber\\ [10pt] & & =\;
\left[
\fnd{\omega^{-3}\;\dissum{i=4}{4}\,\mu_{i}}
{\mu_{1}\mu_{2}\mu_{3}\mu_{4}}
\right]^{3/2}
d^{3}\bm{\rho}_{12}\, d^{3}\bm{\rho}_{34}\, d^{3}\bm{\rho}_{1234}
\; =\;
\left[
\fnd{\omega^{-3}\;\dissum{i=4}{4}\,\mu_{i}}
{\mu_{1}\mu_{2}\mu_{3}\mu_{4}}
\right]^{3/2}
d^{3}\bm{\rho}_{14}\, d^{3}\bm{\rho}_{32}\, d^{3}\bm{\rho}_{1432}
\;\;\; .
\label{wavenorm}
\end{eqnarray}
The second equality in (\ref{wavenorm}) follows because
the transformation $\alpha$ is orthogonal.

Properly normalized solutions for a given eigenvalue $E$ of
(\ref{4pHamiltonian1432}) now take the form
\begin{equation}
\psi^{E}_{\left\{ n,\ell ,m\right\}}
\left(\bm{\rho}_{14}\, ,\;\bm{\rho}_{32}\, ,\;\bm{\rho}_{1432}\right)
\; =\;
\left[
\fnd{\omega^{-3}\;\dissum{i=4}{4}\,\mu_{i}}
{\mu_{1}\mu_{2}\mu_{3}\mu_{4}}
\right]^{-3/4}
\phi_{n_{1},\ell_{1},m_{1}}\left(\bm{\rho}_{14}\right)\,
\phi_{n_{2},\ell_{2},m_{2}}\left(\bm{\rho}_{32}\right)\,
\phi_{n_{3},\ell_{3},m_{3}}\left(\bm{\rho}_{1432}\right)
\;\;\; ,
\label{wave1432}
\end{equation}
where $\left\{ n,\ell ,m\right\}$ stands for the set of quantum numbers
$n_{1}$, $\ell_{1}$, $m_{1}$, $n_{2}$, $\ell_{2}$, $m_{2}$, $n_{3}$, $\ell_{3}$
and $m_{3}\,$,
where
\begin{equation}
E\; =\;\sum_{i=1}^{3}\;\omega\,\left( 2n_{i}+\ell_{i}+\frac{3}{2}\right)
\;\;\; ,
\label{totalE}
\end{equation}
and where the harmonic oscillator wave functions are defined by
\begin{equation}
\phi_{n,\ell ,m}\left(\bm{r}\right)\; =\;
\left[\fnd{2\Gamma\left( n+\ell +\frac{3}{2}\right)}
{\Gamma (n+1)\Gamma\left(\ell +\frac{3}{2}\right)^{2}}\right]^{1/2}
r^{\ell}Y_{m}^{\ell}\left(\hat{r}\,\right)
e^\x{-\frac{1}{2}r^{2}}
{_{1}F_{1}}\left( -n;\,\ell+\frac{3}{2};\, r^{2}\right)
\;\;\; .
\label{HOwavefu}
\end{equation}
Analogously, we have a normalized basis for the solutions of equation
(\ref{4pHamiltonian1234}):
\begin{equation}
\chie^{E}_{\left\{ n',{\ell}',m'\right\}}
\left(\bm{\rho}_{12}\, ,\;\bm{\rho}_{34}\, ,\;\bm{\rho}_{1234}\right)
\; =\;
\left[
\fnd{\omega^{-3}\;\dissum{i=4}{4}\,\mu_{i}}
{\mu_{1}\mu_{2}\mu_{3}\mu_{4}}
\right]^{-3/4}
\phi_{n_{1}',\ell_{1}',m_{1}'}\left(\bm{\rho}_{12}\right)\,
\phi_{n_{2}',\ell_{2}',m_{2}'}\left(\bm{\rho}_{34}\right)\,
\phi_{n_{3}',\ell_{3}',m_{3}'}\left(\bm{\rho}_{1234}\right)
\;\;\; .
\label{wave1234}
\end{equation}
The decay of a meson to a well-defined meson-meson final state,
is now described by matrix elements which determine
the decomposition of the $\psi$-solutions (\ref{wave1432})
on the basis of $\chie$-solutions (\ref{wave1234}).
In this way we translated part of the problem of finding
a transition potential which describes decay,
into the problem of finding decomposition matrix elements
$\cal D$ for an orthogonal transformation $\alpha$:
\begin{eqnarray}
\lefteqn{\psi^{E}_{\left\{ n,\ell ,m\right\}}
\left(\bm{\rho}_{14}\, ,\;\bm{\rho}_{32}\, ,\;\bm{\rho}_{1432}\right)
\; =}
\label{Decompdef}\\ [10pt] & &
\sum_{\left\{ n',{\ell}',m'\right\}}
{\cal D}^{E}_{\left\{
\left\{ n,\ell ,m\right\}\, ,\,\left\{ n',{\ell}',m'\right\}\right\}}
\left(\bm{\rho}_{14}\, ,\;\bm{\rho}_{32}\, ,\;\bm{\rho}_{1432}\, ;\,
\bm{\rho}_{12}\, ,\;\bm{\rho}_{34}\, ,\;\bm{\rho}_{1234}\right)\,
\chie^{E}_{\left\{ n',{\ell}',m'\right\}}
\left(\bm{\rho}_{12}\, ,\;\bm{\rho}_{34}\, ,\;\bm{\rho}_{1234}\right)
\;\;\; .
\nonumber
\end{eqnarray}
Depending on the specific situation under consideration,
one can always find such an orthogonal transformation
which describes the recombination of any system of quarks and antiquarks.
\clearpage

\section{The decay matrix elements}
\label{DME}

In this section we assume that the independent motion in
the centre of mass system of $N+1$ spinless particles,
which move under the influence of harmonic oscillator forces,
can be described by means of $N$ independent and dimensionless
coordinates \bm{r}$_{i}$ and momenta \bm{p}$_{i}$ ($i=$ 1,...,$N$).

Before the recombination, the Hamiltonian of the system reads
\begin{equation}
H\; =\;\sum_{i=1}^{N}\,
\frac{1}{2}\left(\bm{p}_{i}^{2}+\bm{r}_{i}^{2}\right)
\;\;\; .
\label{HamiltonianBefore}
\end{equation}
Furthermore we assume that the recombination of the system can be
described by an orthogonal transformation (repeated indices imply summation)
\begin{equation}
{\bm{r}'}_{i}\; =\;\alpha_{ij}\bm{r}_{j}
\;\;\;\xrm{and}\;\;\;
{\bm{p}'}_{i}\; =\;\alpha_{ij}\bm{p}_{j}
\;\;\; .
\label{alpha}
\end{equation}
After the recombination the Hamiltonian (\ref{HamiltonianBefore})
obtains the form
\begin{equation}
H\; =\;\sum_{i=1}^{N}\,
\frac{1}{2}\left({\bm{p}'}_{i}^{2}+{\bm{r}'}_{i}^{2}\right)
\;\;\; .
\label{HamiltonianAfter}
\end{equation}
Solutions of (\ref{HamiltonianBefore}) are given by
\begin{equation}
\psi^{E}_{\left\{ n,\ell ,m\right\}}
\left(\left\{\bm{r}\right\}\right)
\; =\;\prod_{i=1}^{N}\,
\phi_{n_{i},\ell_{i},m_{i}}\left(\bm{r}_{i}\right)
\;\;\; ,
\label{waveBefore}
\end{equation}
where $\left\{ n,\ell ,m\right\}$ stands for the set of quantum numbers
$n_{1}$, $\ell_{1}$, $m_{1}$, $\dots$, $n_{N}$, $\ell_{N}$, $m_{N}\,$ and
$\left\{\bm{r}\right\}$ for $\bm{r}_{1}$, $\dots$, $\bm{r}_{N}$.
The relation between $\left\{ n,\ell ,m\right\}$ and $E$ reads
\begin{equation}
E\; =\;\sum_{i=1}^{N}\;\left( 2n_{i}+\ell_{i}+\frac{3}{2}\right)
\;\;\; .
\label{totalEB}
\end{equation}
Analogously, solutions of (\ref{HamiltonianAfter}) are of the form
\begin{equation}
\chie^{E}_{\left\{ n',\ell ',m'\right\}}
\left(\left\{\bm{r}'\right\}\right)
\; =\;\prod_{i=1}^{N}\,
\phi_{n_{i}',\ell_{i}',m_{i}'}\left(\bm{r}_{i}'\right)
\;\;\; ,
\label{waveAfter}
\end{equation}
where
\begin{equation}
E\; =\;\sum_{i=1}^{N}\;\left( 2n_{i}'+\ell_{i}'+\frac{3}{2}\right)
\;\;\; .
\label{totalEA}
\end{equation}
Now, we want to generalize the $\cal D$'s of (\ref{Decompdef})
to $N$ oscillators:
\begin{equation}
{\cal D}^{E}_{\left\{
\left\{ n,\ell ,m\right\}\, ,\,\left\{ n',{\ell}',m'\right\}\right\}}
\left(\left\{\bm{r}\right\}\, ;\,\left\{\bm{r}'\right\}\right)
\; =\;
\bracket{\chie^{E}_{\left\{ n',\ell ',m'\right\}}
\left(\left\{\bm{r}'\right\}\right)}
{\psi^{E}_{\left\{ n,\ell ,m\right\}}
\left(\left\{\bm{r}\right\}\right)}
\;\;\; .
\label{Ddef}
\end{equation}
Thereto we introduce the generating function
of harmonic oscillator wave functions
\begin{equation}
G\left(\bm{s}\, ,\,\bm{r}\right)\; =\;
e^\x{-\bm{s}^{2}+2\bm{s}\cdot\bm{r}-\frac{1}{2}\bm{r}^{2}}
\;\;\; .
\label{generatorHO}
\end{equation}
In \cite{PRD25p2406} it has been shown that
\begin{equation}
G\left(\bm{s}\, ,\,\bm{r}\right)\; =\;
\sum_{n,\ell ,m}\beta (n,\ell )\,
s^{2n+\ell}\,
{Y^{\ell}_{m}}^{\ast}\left(\hat{s}\right)\,
\phi_{n,\ell ,m}\left(\bm{r}\right)
\;\;\; ,
\label{generatorHOexpansion}
\end{equation}
where
\begin{displaymath}
\beta (n,\ell )\; =\; (-1)^\x{n}\,
\left[\fnd{2\pi^{3}}
{\Gamma (n+1)\Gamma\left( n+\ell +\frac{3}{2}\right)}\right]^{1/2}
\;\;\; .
\end{displaymath}

We first notice that for the orthogonal transformation
(\ref{alpha}) follows
\begin{equation}
\prod_{i=1}^{N}\,
G\left(\bm{s}_{i}\, ,\,\bm{r}_{i}'\right)\; =\;
\prod_{i=1}^{N}\,
G\left(\bm{s}_{i}\, ,\,\alpha_{ij}\bm{r}_{j}\right)\; =\;
\left[
\prod_{i=1}^{N}\,\prod_{j=1}^{N}\,
G\left(\alpha_{ij}\bm{s}_{i}\, ,\,\bm{r}_{j}\right)
\right]\,
e^\x{\frac{1}{2}(N-1)\dissum{k=1}{N}\,\bm{r}_{k}^{2}}
\;\;\; .
\label{Galpha}
\end{equation}
If we then multiply with the product
$Y^{\ell_{1}'}_{m_{1}'}\left(\hat{s}_{1}\right)\,\dots\,
Y^{\ell_{N}'}_{m_{N}'}\left(\hat{s}_{N}\right)$,
integrate both sides of Eq.~(\ref{Galpha}) over the volume
$d\Omega_{\bm{s}_{1}}\,\dots\, d\Omega_{\bm{s}_{N}}$
and equate the coefficients
of the equal powers of $s_{1}$, $\dots$, $s_{N}$,
then we find for (\ref{waveAfter}) the expression
\begin{eqnarray}
\lefteqn{\chie^{E}_{\left\{ n',\ell ',m'\right\}}
\left(\left\{\bm{r}'\right\}\right)
\; =}
\label{HOexpansion}\\ [10pt] & & =\;
(4\pi )^{\frac{1}{2}N(N-1)}\,
e^\x{\frac{1}{2}(N-1)\dissum{k=1}{N}\,\bm{r}_{k}^{2}}\,
\prod_{i=1}^{N}\,\beta\left( n_{i}',\ell_{i}'\right)^{-1}
\sum_{n_{ij},\ell_{ij},m_{ij}}\,
\delta\left(\sum_{j=1}^{N}\left( 2n_{ij}+\ell_{ij}\right)\, ,\,
2n_{i}'+\ell_{i}'\right)\,\times
\nonumber\\ [10pt] & & \times\,
\left[
\fnd{\disprod{j=1}{N}\,\left( 2\ell_{ij}+1\right)}{2\ell_{i}'+1}
\right]^{1/2}
\left(\begin{array}{ccc}
\ell_{i1} & \cdots & \ell_{iN}\\ [10pt]
m_{i1} & \cdots & m_{iN}\end{array}\right|\left.
\begin{array}{c}\ell_{i}'\\ [10pt] m_{i}'\end{array}\right)\,
\prod_{j=1}^{N}\,
\beta\left( n_{ij},\ell_{ij}\right)\,
\left(\alpha_{ij}\right)^{2n_{ij}+\ell_{ij}}\,
\phi_{n_{ij},\ell_{ij},m_{ij}}\left(\bm{r}_{j}\right)
\;\;\; .
\nonumber
\end{eqnarray}
Here $\left\{ n_{ij},\ell_{ij},m_{ij}\right\}$
stands for the set of 3 times $N\times N$ quantum numbers
$n_{11},\ell_{11},m_{11}$, $\dots$, $n_{NN},\ell_{NN},m_{NN}$.
An exception occurs if one (or more) of the matrix elements
$\alpha_{ij}$ equals zero.
In this case it is sufficient to replace
$\left(\alpha_{ij}\right)^{2n_{ij}+\ell_{ij}}$
by $\delta_{n_{ij},0}\delta_{\ell_{ij},0}$,
which replacement can be understood
if one puts $s=0$ in (\ref{generatorHOexpansion}).
The symbol between brackets in (\ref{HOexpansion}) is defined by
\begin{eqnarray*}
\lefteqn{
\left(\begin{array}{ccc}
\ell_{i1} & \cdots & \ell_{iN}\\ [10pt]
m_{i1} & \cdots & m_{iN}\end{array}\right|\left.
\begin{array}{c}\ell_{i}'\\ [10pt] m_{i}'\end{array}\right)\; =}
\\ [10pt] & & =\;
(4\pi )^{\frac{1}{2}(N-1)}\,
\left[
\fnd{2\ell_{i}'+1}{\disprod{j=1}{N}\,\left( 2\ell_{ij}+1\right)}
\right]^{1/2}
\int d\Omega_{\bm{s}_{i}}\,
{Y^{\ell_{i1}}_{m_{i1}}}^{\ast}\left(\hat{s}_{i}\right)\,\dots\,
{Y^{\ell_{iN}}_{m_{iN}}}^{\ast}\left(\hat{s}_{i}\right)\,
Y^{\ell_{i}'}_{m_{i}'}\left(\hat{s}_{i}\right)
\;\;\; .
\end{eqnarray*}
With \cite{Rose} this can be written as a sum of products
of $SO(3)$-Clebsch-Gordan coefficients:
\begin{eqnarray}
\lefteqn{
\left(\begin{array}{ccc}
\ell_{1} & \cdots & \ell_{N}\\ [10pt]
m_{1} & \cdots & m_{N}\end{array}\right|\left.
\begin{array}{c}\ell\\ [10pt] m\end{array}\right)\; =}
\label{RCGgeneralized}\\ [10pt] & & =\;
\sum_{\left\{ L_{i},M_{i}\right\}}
\delta_{L_{1},\ell_{1}}\,
\delta_{M_{1},m_{1}}\,
\delta_{L_{N},\ell}\,
\delta_{M_{N},m}\,
\prod_{i=1}^{N-1}\,
\Clebsch{L_{i}}{\ell_{i+1}}{L_{i+1}}{M_{i}}{m_{i+1}}{M_{i+1}}
\Clebsch{L_{i}}{\ell_{i+1}}{L_{i+1}}{0}{0}{0}
\;\;\; .
\nonumber
\end{eqnarray}
The matrix element (\ref{Ddef}) can be written in the form
(using (\ref{HOexpansion}), (\ref{RCGgeneralized})
and the result of the Appendix~\ref{append})
\begin{eqnarray}
\lefteqn{{\cal D}^{E}_{\left\{
\left\{ n,\ell ,m\right\}\, ,\,\left\{ n',{\ell}',m'\right\}\right\}}
\left(\left\{\bm{r}\right\}\, ;\,\left\{\bm{r}'\right\}\right)
\; =\;
\left(\fnd{\pi}{4}\right)^{\frac{1}{2}N(N-1)}
}
\nonumber\\ [10pt] & & \!\!\!\!\!
\left\{
\prod_{i=1}^{N}(-1)^{n_{i}}
\left[\fnd{\Gamma\left(n_{i}+1\right)
\Gamma\left(n_{i}+\ell_{i}+\frac{3}{2}\right)}{2\ell_{i}+1}\right]^{1/2}
\right\}
\left\{
\prod_{j=1}^{N}(-1)^{n_{j}'}
\left[\fnd{\Gamma\left(n_{j}'+1\right)
\Gamma\left(n_{j}'+\ell_{j}'+\frac{3}{2}\right)}{2\ell_{j}'+1}\right]^{1/2}
\right\}
\nonumber\\ [10pt] & & \!\!\!\!\!
\sum_{n_{ij},\ell_{ij},m_{ij}}\;\;
\left[\;\;
\left\{
\prod_{i=1}^{N}\,
\delta\left(\sum_{j=1}^{N}\left( 2n_{ij}+\ell_{ij}\right)\, ,\,
2n_{i}'+\ell_{i}'\right)\,
\left(\begin{array}{ccc}
\ell_{i1} & \cdots & \ell_{iN}\\ [10pt]
m_{i1} & \cdots & m_{iN}\end{array}\right|\left.
\begin{array}{c}\ell_{i}'\\ [10pt] m_{i}'\end{array}\right)
\right\}
\right.
\nonumber\\ [10pt] & & \;\;\;\;\;\;\;\;\;\;\;\;\;\;\;\;
\left\{
\prod_{j=1}^{N}\,
\delta\left(\sum_{i=1}^{N}\left( 2n_{ij}+\ell_{ij}\right)\, ,\,
2n_{j}+\ell_{j}\right)\,
\left(\begin{array}{ccc}
\ell_{1j} & \cdots & \ell_{Nj}\\ [10pt]
m_{1j} & \cdots & m_{Nj}\end{array}\right|\left.
\begin{array}{c}\ell_{j}\\ [10pt] m_{j}\end{array}\right)
\right\}
\nonumber\\ [10pt] & & \left.
\left\{
\prod_{i=1}^{N}\,\prod_{j=1}^{N}\,
\left(\alpha_{ij}\right)^{2n_{ij}+\ell_{ij}}\,
\fnd{2\ell_{ij}+1}{\Gamma\left(n_{ij}+1\right)
\Gamma\left(n_{ij}+\ell_{ij}+\frac{3}{2}\right)}
\right\}
\;\;\right]
\;\;\; .
\label{Dfinal}
\end{eqnarray}

This expression is the central result of this paper.

\begin{figure}[htbp]
\begin{center}
\begin{tabular}{c}
\includegraphics[height=400pt]
{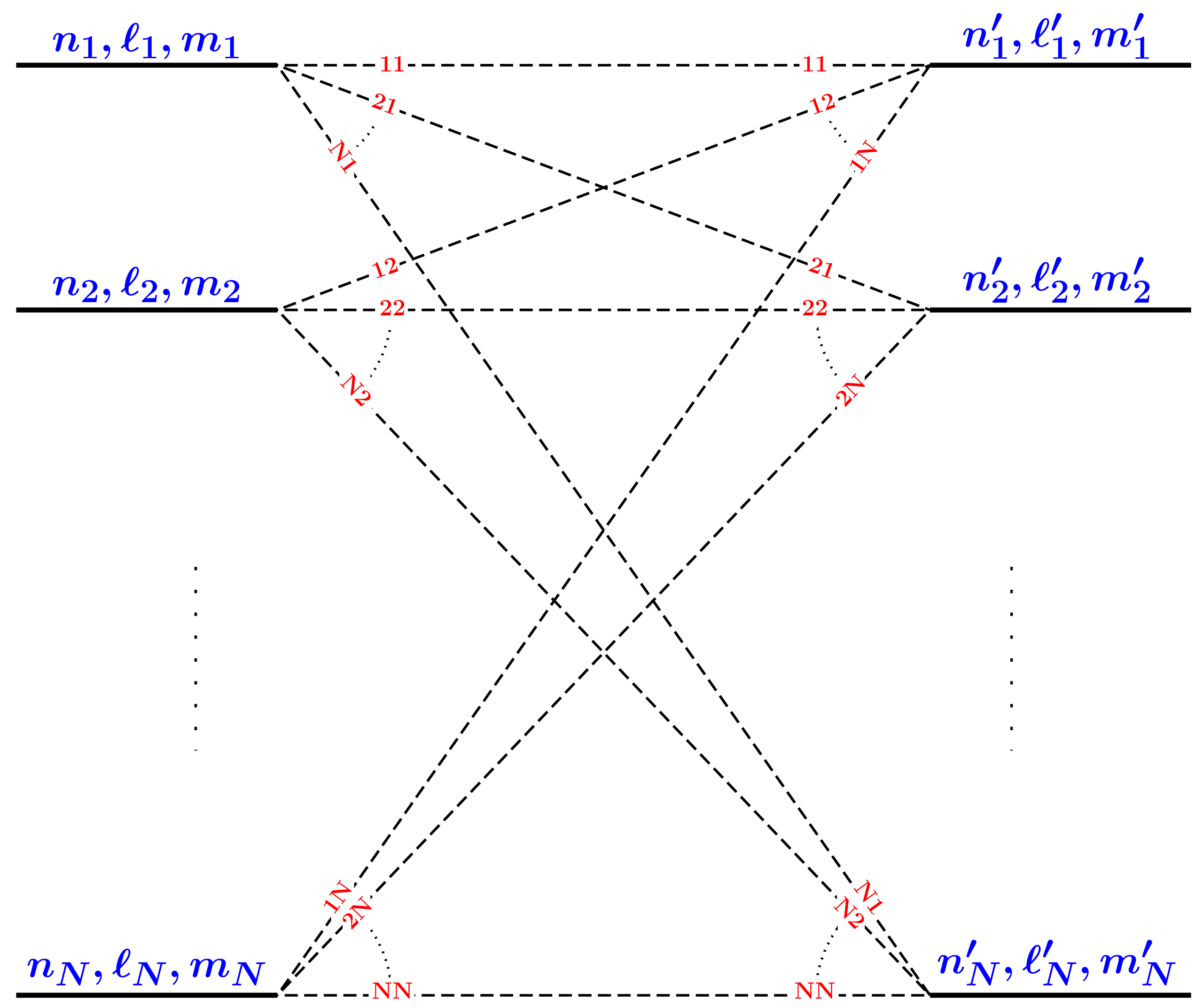}
\end{tabular}
\end{center}
\caption[]{\small Diagrammatic representation of formula (\ref{Dfinal}).}
\label{diagrep}
\end{figure}

With an extension of the rules defined in Ref.~\cite{PRD25p2406},
formula (\ref{Dfinal}) can be represented by a diagram as depicted in
Fig.~\ref{diagrep}.
Here, every left vertex $\left\{ n_{j},\ell_{j},m_{j}\right\}$,
where the internal lines, $\left( n_{1j},\ell_{1j},m_{1j}\right)$,
$\dots$, $\left( n_{Nj},\ell_{Nj},m_{Nj}\right)$, end,
contributes a factor
\begin{eqnarray}
\lefteqn{(-1)^{n_{j}}\,
\left(\fnd{\pi}{4}\right)^{-\frac{1}{4}}\,
\left[\fnd{\Gamma\left(n_{j}+1\right)
\Gamma\left(n_{j}+\ell_{j}+\frac{3}{2}\right)}{2\ell_{j}+1}\right]^{1/2}
\,\times}
\nonumber\\ [10pt] & & \times\,
\left(\begin{array}{ccc}
\ell_{1j} & \cdots & \ell_{Nj}\\ [10pt]
m_{1j} & \cdots & m_{Nj}\end{array}\right|\left.
\begin{array}{c}\ell_{j}\\ [10pt] m_{j}\end{array}\right)\,
\delta\left(\sum_{i=1}^{N}\left( 2n_{ij}+\ell_{ij}\right)\, ,\,
2n_{j}+\ell_{j}\right)
\;\;\; ,
\label{left_vertex}
\end{eqnarray}
every rightvertex, $\left\{ n_{i}',\ell_{i}',m_{i}'\right\}$,
where the internal lines, $\left( n_{i1},\ell_{i1},m_{i1}\right)$,
$\dots$, $\left( n_{iN},\ell_{iN},m_{iN}\right)$, end,
contributes a factor
\begin{eqnarray}
\lefteqn{(-1)^{n_{i}'}\,
\left(\fnd{\pi}{4}\right)^{-\frac{1}{4}}\,
\left[\fnd{\Gamma\left(n_{i}'+1\right)
\Gamma\left(n_{i}'+\ell_{i}'+\frac{3}{2}\right)}{2\ell_{i}'+1}\right]^{1/2}
\,\times}
\nonumber\\ [10pt] & & \times\,
\left(\begin{array}{ccc}
\ell_{i1} & \cdots & \ell_{iN}\\ [10pt]
m_{i1} & \cdots & m_{iN}\end{array}\right|\left.
\begin{array}{c}\ell_{i}'\\ [10pt] m_{i}'\end{array}\right)\,
\delta\left(\sum_{j=1}^{N}\left( 2n_{ij}+\ell_{ij}\right)\, ,\,
2n_{i}'+\ell_{i}'\right)
\;\;\; ,
\label{right_vertex}
\end{eqnarray}
and every internal line, $\left\{ n_{ij},\ell_{ij},m_{ij}\right\}$,
contributes a factor
\begin{equation}
\left(\fnd{\pi}{4}\right)^{\frac{1}{2}}\,
\left(\alpha_{ij}\right)^{2n_{ij}+\ell_{ij}}\,
\fnd{2\ell_{ij}+1}{\Gamma\left(n_{ij}+1\right)
\Gamma\left(n_{ij}+\ell_{ij}+\frac{3}{2}\right)}
\;\;\; .
\label{internal_line}
\end{equation}
\clearpage

\section{The transition potential for the decay \bm{\rho\to\pi\pi}}
\label{MtoMMTpot}

If, just before the decay of a $\rho$ meson,
a $^{3}P_{0}$ pair has been created,
we are in the situation in which two particles
(coming from the $\rho$ meson) are in a relative $S$ wave,
in which the $^{3}P_{0}$ pair is obviously in a relative P-wave
and in which as we shall assume,
both pairs are in a relative $S$ wave.
Just after the decay of the $\rho$ meson there are two pairs (the pions).
The particles inside each of the pions are in relative $S$ waves,
and the pions are in a relative $P$ wave.
\begin{figure}[htbp]
\begin{center}
\begin{tabular}{c}
\includegraphics[height=200pt]
{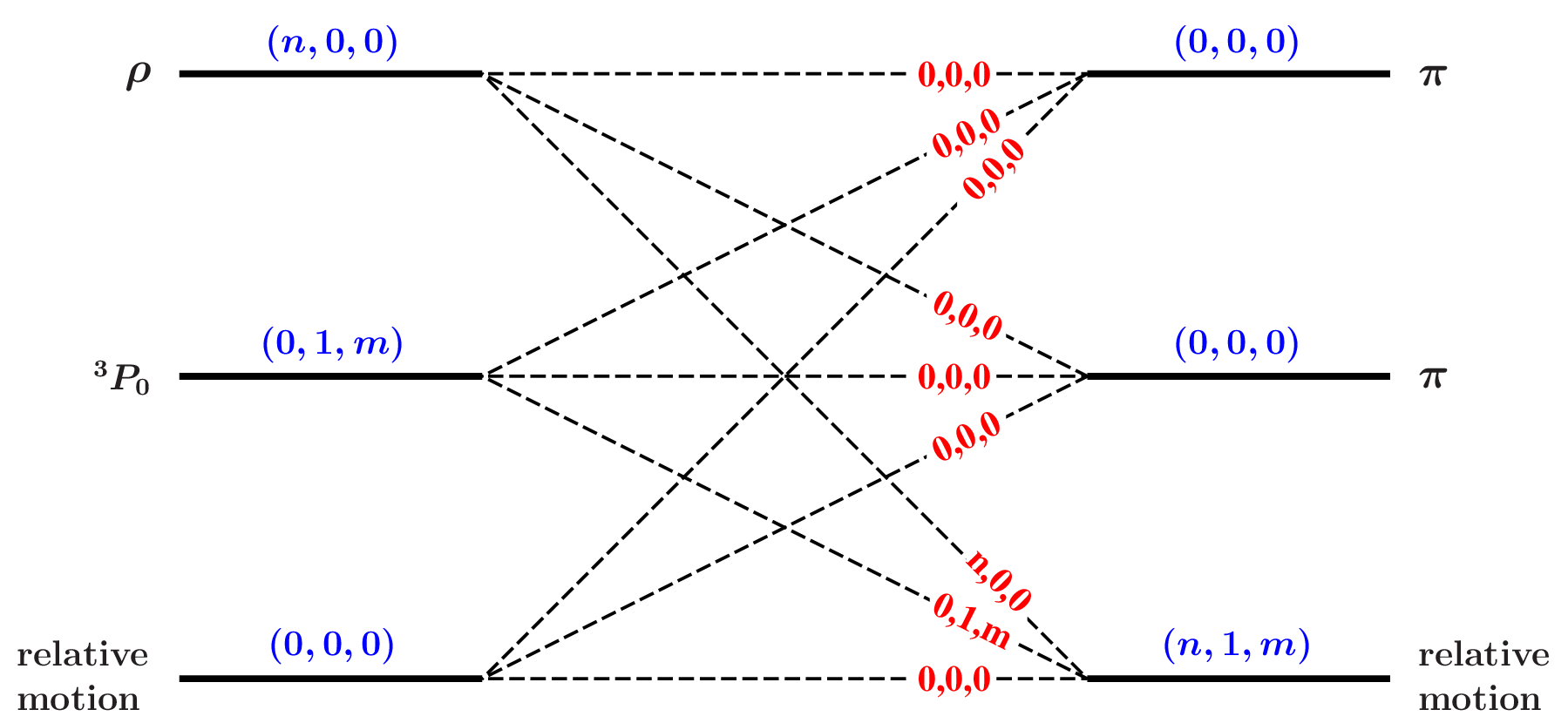}
\end{tabular}
\end{center}
\caption[]{\small Diagram for $\rho\to\pi\pi$.}
\label{rhotopipi}
\end{figure}
The diagram for this rearrangement is shown in Fig.~\ref{rhotopipi},
from which one may conclude that we only allow
radial excitations in the $\rho$ meson initial state and
in the relative final-state motion of the two pions.

Because we assume equal masses for the participating quarks,
the orthogonal transformation (\ref{alpha}) takes the form:
\begin{equation}
\alpha\; =\;
\left(\begin{array}{ccc}
\frac{1}{2} & \frac{1}{2} &  \sqrt{\frac{1}{2}}\\ [10pt]
\frac{1}{2} & \frac{1}{2} & -\sqrt{\frac{1}{2}}\\ [10pt]
\sqrt{\frac{1}{2}} & -\sqrt{\frac{1}{2}} & 0\end{array}\right)
\;\;\; .
\label{rhotopipitrafo}
\end{equation}
There is only one non-vanishing contribution to
the sum of possibilities of the internal lines
in the diagram of Fig.~\ref{rhotopipi} (this possibility is shown!).

For the calculation of the diagram we need the following vertices
\begin{eqnarray}
& &
\begin{tabular}{c}
\includegraphics[height=100pt]
{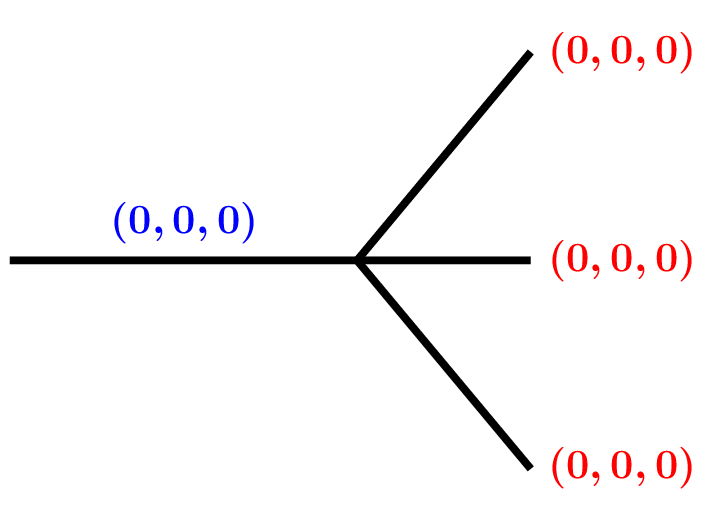}
\end{tabular}
\; =\;
\begin{tabular}{c}
\includegraphics[height=100pt]
{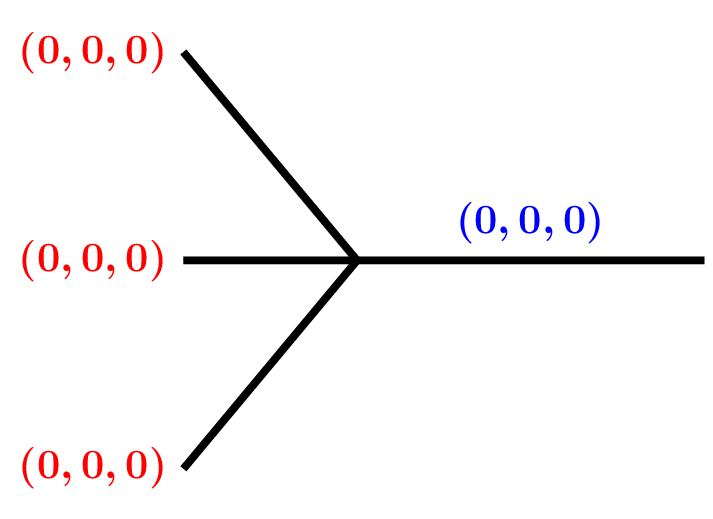}
\end{tabular}
\; =\; 1
\;\;\; ,
\nonumber\\ [10pt] & &
\begin{tabular}{c}
\includegraphics[height=100pt]
{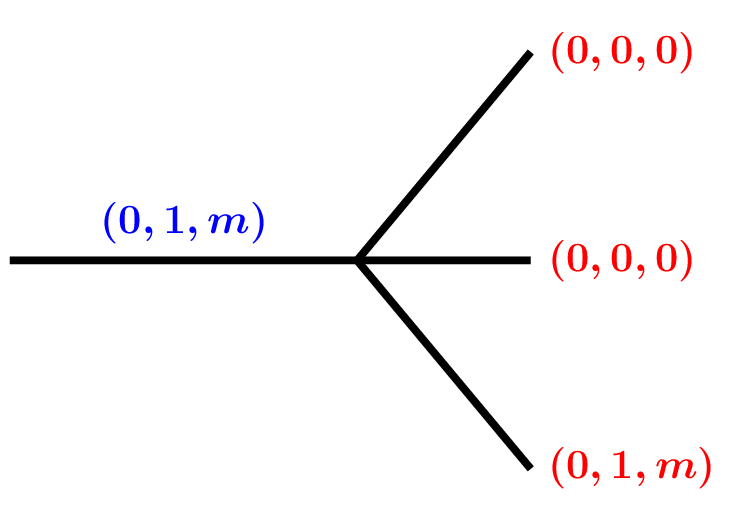}
\end{tabular}
\; =\;\sqrt{\frac{1}{2}}
\;\;\; ,
\nonumber\\ [10pt] & &
\begin{tabular}{c}
\includegraphics[height=100pt]
{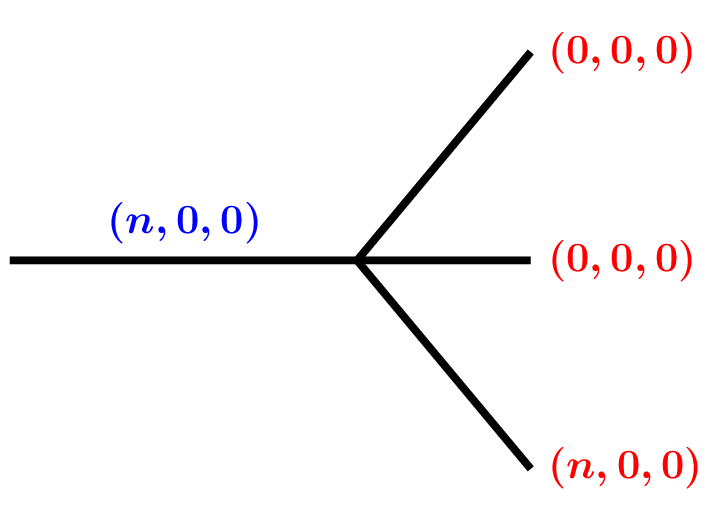}
\end{tabular}
\; =\;
(-1)^{n}\,
\left(\fnd{\pi}{4}\right)^{-\frac{1}{4}}\,
\sqrt{\Gamma\left(n+1\right)\Gamma\left(n+\frac{3}{2}\right)}
\;\;\; ,
\nonumber\\ [10pt] & &
\begin{tabular}{c}
\includegraphics[height=100pt]
{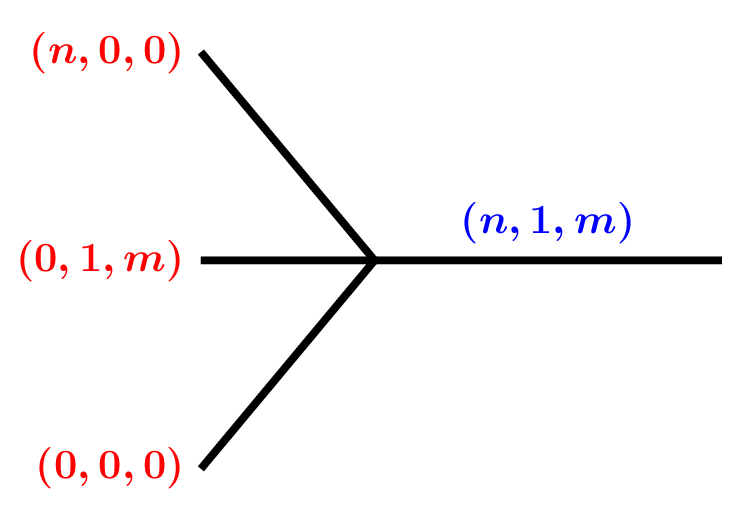}
\end{tabular}
\; =\;
(-1)^{n}\,
\left(\fnd{\pi}{4}\right)^{-\frac{1}{4}}\,
\sqrt{\fnd{\Gamma\left(n+1\right)
\Gamma\left(n+\fndrs{2pt}{5}{-4pt}{2}\right)}{3}}
\;\;\; .
\label{vertices}
\end{eqnarray}
and the internal lines
\begin{eqnarray}
& &
\begin{tabular}{c}
\includegraphics[height=25pt]
{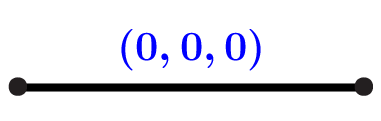}
\end{tabular}
\; =\; 1
\;\;\; ,
\nonumber\\ [10pt] & &
\begin{tabular}{c}
\includegraphics[height=25pt]
{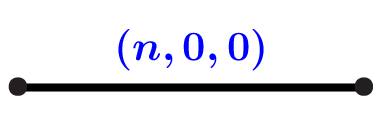}
\end{tabular}
\; =\;
\left(\fnd{\pi}{4}\right)^{\frac{1}{2}}\,
\fnd{1}{\Gamma\left(n+1\right)\Gamma\left(n+\frac{3}{2}\right)}
\;\;\; ,
\nonumber\\ [10pt] & &
\begin{tabular}{c}
\includegraphics[height=25pt]
{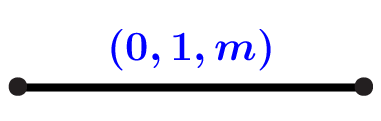}
\end{tabular}
\; =\; -\sqrt{2}
\;\;\; .
\label{internal}
\end{eqnarray}
With the use of formulae (\ref{vertices}) and (\ref{internal}),
we obtain for the diagram the result:
\begin{equation}
-\left(\frac{1}{2}\right)^{n}\,
\sqrt{\fnd{2n+3}{6}}
\;\;\; .
\label{Drhotopipi}
\end{equation}

The result (\ref{Drhotopipi}) can be used for the calculation
of a transition potential.
In a specific model we consider the quarks in the
$\rho$ meson as pointlike particles.
The initial state is thus described by a wave function
$\bracket{\phi}{\bm{x}}$, where \bm{x} is the separation between the quarks.

Similarly we treat the final-state pions as pointlike objects
for which we take the same spatial separation as for
the quark and antiquark in the initial state.
The rearrangement potential which follows from this treatment equals
\begin{equation}
\braket{\bm{x}}{V}{\bm{x}'}\; =\;
\sum_{n,n'}\,\bracket{\bm{x}}{n}
\braket{n}{V}{n'}\bracket{n'}{\bm{x}'}
\;\;\; .
\label{potential}
\end{equation}
In (\ref{potential}) we have inserted a complete set
of harmonic oscillator wave functions
\begin{equation}
\bracket{\bm{x}}{n}\;\propto\;
e^\x{-\frac{1}{2}\fnd{m}{2}\omega x^{2}}\,
\phi\left( -n\, ;\,\frac{3}{2}\, ;\,\fnd{m}{2}\omega x^{2}\right)
\;\;\; .
\label{rhowavefunction}
\end{equation}
because the $\rho$ meson is in an $S$ wave,
\begin{equation}
\bracket{n'}{\bm{x}'}\;\propto\;
x'\, e^\x{-\frac{1}{2}m\omega {x'}^{2}}\,
\phi\left( -n'\, ;\,\frac{5}{2}\, ;\, m\omega {x'}^{2}\right)
\;\;\; .
\label{pipiwavefunction}
\end{equation}
because the pions are in a relative $P$ wave.
The masses in (\ref{rhowavefunction}) and (\ref{pipiwavefunction})
are such that a pion is
twice as heavy as a (constituent) quark.

If we insert in (\ref{potential}) for $\braket{n}{V}{n'}$
the diagram of Fig.~\ref{rhotopipi},
we clearly see that $n'=n$.
Using (\ref{Drhotopipi}), (\ref{rhowavefunction}) and
(\ref{pipiwavefunction}), we obtain for (\ref{potential}) the result
\begin{equation}
\braket{\bm{x}}{V}{\bm{x}'}\;\propto\;
\left(\frac{1}{2}\right)^{n}\,
\sqrt{\fnd{2n+3}{6}}\,
x'\,
e^\x{-\frac{1}{2}m\omega\left(\frac{1}{2}x^{2}+{x'}^{2}\right)}\,
\phi\left( -n\, ;\,\frac{3}{2}\, ;\,\fnd{m}{2}\omega x^{2}\right)
\phi\left( -n\, ;\,\frac{5}{2}\, ;\, m\omega {x'}^{2}\right)
\;\;\; .
\label{nonlocal}
\end{equation}
It is common practice to approximate this by a local one:
\begin{displaymath}
V(x)\; =\;\braket{\bm{x}}{V}{\bm{x}'}\,
\delta\left(x'-x\right)
\end{displaymath}
in this case we obtain
\begin{equation}
V(x)\;\propto\;
\left(\frac{1}{2}\right)^{n}\,
\sqrt{\fnd{2n+3}{6}}\,
x\,
e^\x{-\frac{3}{4}m\omega x^{2}}\,
\phi\left( -n\, ;\,\frac{3}{2}\, ;\,\fnd{m}{2}\omega x^{2}\right)
\phi\left( -n\, ;\,\frac{5}{2}\, ;\, m\omega x^{2}\right)
\;\;\; .
\label{local}
\end{equation}
The form of the potential (\ref{local}) is in agreement with
the phenomenological potential of \cite{PRD21p772}
\begin{equation}
V(x)\;\propto\; x\, e^\x{-\beta m\omega x^{2}}
\;\;\; .
\label{phenomenological}
\end{equation}
where the authors quote $\beta=\xrm{(about) }1.7$.
\clearpage

\section{Conclusions}

The advantage of the use of harmonic oscillators
for the calculation of strong decay potentials is not only
that the calculations are very easy and that the number
of possible final states is limited,
but also that it is possible to determine
the spatial dependence of the strong decay potential.
Whether or not this spatial dependence is in agreement
with experiment,
might be checked within the decay model proposed in
\cite{AP105p318,PRD21p772}.
However, since the position of the maximum of (\ref{local})
is in accordance with the maximum of (\ref{phenomenological}),
not much disagreement has to be expected.
\vspace{.3cm}

{\large\bf Acknowledgments}
\vspace{.3cm}

I am grateful to Dr. J.E. Ribeiro of the Centro Fisica Materia Condensada,
Lisbon (Portugal) for suggesting this problem
and taking part in the formulation of the program to be followed.
Furthermore I would like to thank Professor Dr. C. Dullemond
and Drs. L. Somers for many useful conversations.

This work is part of the research program of the Stichting
voor Fundamenteel Onderzoek der Materie (F.O.M.)
with financial support from the Nederlandse Organizatie
voor ZuiverWetenschappelijk Onderzoek (Z.W.O.).
\clearpage

\appendix

\section{Appendix}
\label{append}

The derivation of (\ref{Dfinal}) involves the following product
of integrations:
\begin{equation}
\prod_{j=1}^{N}\,\int\, d^{3}r_{j}\,
\left[\,
\prod_{i=1}^{N}\,\phi_{n_{ij},\ell_{ij},m_{ij}}\left(\bm{r}_{j}\right)
\right]\,
\phi^{\ast}_{n_{j},\ell_{j},m_{j}}\left(\bm{r}_{j}\right)\,
e^\x{-\frac{1}{2}\left( N-1\right)\,\bm{r}_{j}^{2}}
\;\;\; ,
\label{A1}
\end{equation}
which equals to (using Eq.~\ref{RCGgeneralized}):
\begin{eqnarray}
& & \prod_{j=1}^{N}\,\left\{\,
\delta\left(\sum_{i=1}^{N}\left( 2n_{ij}+\ell_{ij}\right)\, ,\,
2n_{j}+\ell_{j}\right)\,
\left(\begin{array}{ccc}
\ell_{1j} & \cdots & \ell_{Nj}\\ [10pt]
m_{1j} & \cdots & m_{Nj}\end{array}\right|\left.
\begin{array}{c}\ell_{j}\\ [10pt] m_{j}\end{array}\right)
\right.
\label{A2}\\ [10pt] & & \!\!\!\!\!\left.
(-1)^{n_{j}}\,\left(\fnd{1}{2\pi}\right)^{\frac{1}{2}(N-1)}
\left[\fnd{\Gamma\left(n_{j}+1\right)
\Gamma\left(n_{j}+\ell_{j}+\frac{3}{2}\right)}{2\ell_{j}+1}\right]^{1/2}\,
\prod_{i=1}^{N}\,(-1)^{n_{ij}}\,
\fnd{2\ell_{ij}+1}{\Gamma\left(n_{ij}+1\right)
\Gamma\left(n_{ij}+\ell_{ij}+\frac{3}{2}\right)}
\right\}\, .
\nonumber
\end{eqnarray}
Because the derivation of the result (\ref{A2}) is rather lengthy,
we will not present its full details,
but only outline a possible procedure.

The delta functions in (\ref{A2}) follow from the fact that the functions
\begin{equation}
e^\x{-\frac{1}{2}\left( N-1\right)\,\bm{r}_{j}^{2}}\,
\prod_{i=1}^{N}\,\phi_{n_{ij},\ell_{ij},m_{ij}}\left(\bm{r}_{j}\right)
\;\;\; ,
\label{A3}
\end{equation}
are eigenfunctions of
$H_{j}=\frac{1}{2}\left(\bm{x}_{j}^{2}+\bm{p}_{j}^{2}\right)$,
with eigenvalues
\begin{displaymath}
\sum_{i=1}^{N}\,\left( 2n_{ij}+\ell_{ij}+\frac{3}{2}\right)
\;\;\; .
\end{displaymath}
From the definition of the harmonic oscillator wavefunctions
(\ref{HOwavefu}),
it can be seen that the integrations over the solid angles
are exactly those already defined in Eq.~(\ref{RCGgeneralized}).
The remaining parts of the integrations may be performed
by introducing the series expansions of the confluent
hypergeometric functions.
This leads to standard integrals:
\begin{eqnarray*}
\lefteqn{
\int_{0}^{\infty}\, r_{j}^{2}\, dr_{j}\,
\left[\,\prod_{i=1}^{N}\,\left( r_{j}\right)^{\ell_{ij}}\,
{_{1}F_{1}}\left( -n_{ij};\,\ell_{ij}+\frac{3}{2};\, r_{j}^{2}\right)
\right]\,
{_{1}F_{1}}\left( -n_{j};\,\ell_{j}+\frac{3}{2};\, r_{j}^{2}\right)\,
e^\x{-\frac{1}{2}r_{j}^{2}}
\; =}
\\ [10pt] & & \!\!\!\!\! =\;
\fnd{\Gamma\left(\ell_{j}+\frac{3}{2}\right)}
{\Gamma\left(n_{j}+\ell_{j}+\frac{3}{2}\right)}\,
\left[\,\prod_{i=1}^{N}\,
\fnd{\Gamma\left(\ell_{ij}+\frac{3}{2}\right)}
{\Gamma\left( -n_{ij}\right)}\right]\,
\sum_{k_{1},\dots ,k_{N}}\,\left\{
\left[\,\prod_{i=1}^{N}\,
\fnd{\Gamma\left( k_{i}-n_{ij}\right)}{\Gamma\left( k_{i}+1\right)
\Gamma\left( k_{i}+\ell_{ij}+\frac{3}{2}\right)}\right]
\right.
\\ [10pt] & & \!\!\!\!\!\left.
\Gamma\left(\sum_{i=1}^{N}\,\left( k_{i}+\frac{1}{2}\ell_{ij}\right)
+\frac{1}{2}\ell_{j}+\frac{3}{2}\right)\,
\fnd
{\Gamma\left( n_{j}+\frac{1}{2}\ell_{j}-
\dissum{i=1}{N}\,\left( k_{i}+\frac{1}{2}\ell_{ij}\right)\right)}
{2\Gamma\left(\frac{1}{2}\ell_{j}-
\dissum{i=1}{N}\,\left( k_{i}+\frac{1}{2}\ell_{ij}\right)\right)}
\,\right\}
\end{eqnarray*}

At this stage one must also use the fact that the angular momenta
are restricted because the $\ell$'s in the symbol defined in
Eq.~(\ref{RCGgeneralized}),
must add up to an even integer.
Of the resulting product of sums only one term of each sum contributes,
which leads finally to the expression (\ref{A2}).

\end{document}